# Life time of correlation between stocks prices on established and emerging markets

## Andrzej Buda[a,b]

[a]Wydawnictwo Niezalezne, ul. Oriona 15/8 67-200 Głogów, Poland
[b]Institute of Theoretical Physics, Wrocław University, 50-204 Wrocław, pl. M. Borna 9, Poland

Abstract

The correlation coefficient between stocks depends on price history and includes information on hierarchical structure in financial markets. It is useful for portfolio selection and estimation of risk. I introduce the Life Time of Correlation between stocks prices to know how far we should investigate the price history to obtain the optimal durability of correlation. I carry out my research on emerging (Poland) and established markets (in the USA, Great Britain and Germany). Other methods, including the Minimum Spanning Trees, tree half-life, decomposition of correlations and the Epps effect are also discussed.

## 1 Introduction

The correlation coefficients between stocks determines the success of a particular diversification strategy [1-2] and reflects hierarchical structure in financial markets [3-4]. It consequently depends upon the quality of the estimated correlation between stocks. It is well known, however, that there is a tendency for the average correlation among stocks to change as markets rise and fall. There has been a lot of research into question of random processes in time evolution of stock returns and correlations, since [5-7] up to [8-11]. Consequently, the constancy of correlation between stocks over time seems unrealistic.

The traditional approach to the estimation of the correlation coefficient between stocks is to use a fixed number of time series observations with a sufficient large number of data points to provide statistically significant estimates [1-2]. Such an estimation method is deficient. The time series method provides only an unconditional estimate and so changes in correlation coefficients are difficult to judge. Even using a "moving window" of overlapping observations (by replacing one month from the beginning of the data series with the latest observation) is unsatisfactory because any two successive correlation coefficients are based on almost identical data sets. Hence, a long time is needed for a noticeable change in the general level of correlation to be reflected in the estimation. While, a temporary change will go unnoticed because it affects only a few observations in the estimation window.

For this purpose, a new tool is proposed, namely lifetime of correlation between stocks prices to investigate the durability of correlations. The research was carried out on established and emerging markets. The results are discussed in Section 4.

## 2 The correlation coefficient

The correlation coefficient defines degree of similarity between the synchronous time evolution of a pair of stocks prices.

$$\rho_{ij} = \frac{\langle Y_i Y_j \rangle - \langle Y_i \rangle \langle Y_j \rangle}{\sqrt{(\langle Y_i^2 \rangle - \langle Y_i \rangle^2)(\langle Y_j^2 \rangle - \langle Y_j \rangle^2)}} \qquad (1)$$

where $i$ and $j$ are the numerical labels of stocks, *Yij* is the price return. *Yi* = ln *Pi*(*t*) - ln *Pi*(*t* – 1) where *Pi*(*t*) is the closure price of the stock $i$ at the day $t$. The statistical average is a temporal average performed on all the trading days of investigated time period. By definition, ρ*ij* may vary from -1 to 1. The matrix of correlation coefficients is a symmetric matrix with ρ*ii*. The

n(n-1)/2 correlation coefficients characterize the matrix completely. It not only reflects economic connections in financial markets, but is also useful to calculate risks of investment. Markovitz concept of Portfolio Selection [1] allows to get the return at the minimum risk possible. It is based on the correlation coefficients that have main importance in strategies of investment.

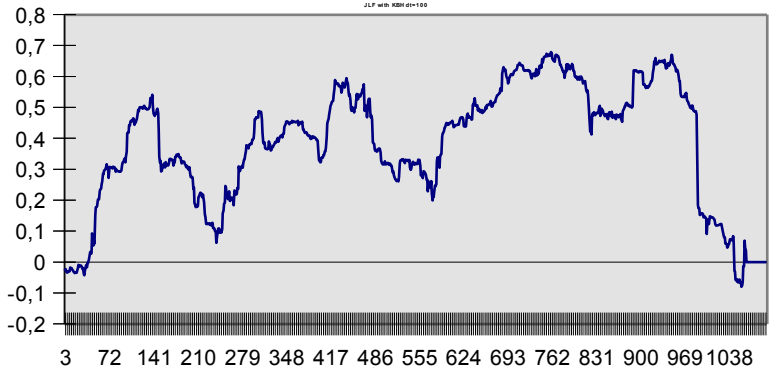

**Fig. 1** The evolution of the correlation between JLF and KBH from The WIG Portfolio (1997-2004) for the window width $\Delta t$ = 100 trading days [t.d.].

Despite its theoretical importance, the Modern Portfolio Theory does not match the real world because correlations between assets depend on systemic relationships between the underlying assets, and vary when these relationships change.

The correlation coefficient reflects similarity between stocks. It can be used in building the hierarchical structure in financial markets and finding the taxonomy that allows to isolate groups of stocks that make sense from an economic point of view [4]. The correlation coefficient cannot be used as a distance between stocks because it does not fulfill the three axioms that define a metric. However a metric can be defined using a distance function of the correlation coefficient.

The introduction of a distance between a synchronous evolving pair of stocks was proposed correctly in 1999 by R.N. Mantegna [3] according to Didier Sornette's suggestion. A metric can be defined using as distance a function of the correlation coefficient:

$$d(i,j) = \sqrt{2\left(1 - \rho_{ij}\right)} \qquad (2)$$

**Fig. 2** Minimum Spanning Tree (MST) connecting the stocks used to compute:
A) Warszawski Indeks Giełdowy (WIG 20)
B) Deutsche Aktienindex (DAX)
C) Dow Jones Industrial Average (DJIA)

D) FTSE 1000

With this choice, d(i,j) fulfills three axioms of an Euclidean metric:

(i)   d **ij** = 0                  if and only if     i = j
(ii)  d **ij** = d **ji**
(iii) d **ij** < d **ik** + d **kj**

In my research I investigate correlation coefficients among stocks traded in established (USA, Germany, Great Britain) and emerging (Poland) markets.

I choose 4 set of stocks used to compute main indices: 30 stocks used to compute the Dow Jones Industrial Average (DJIA) 26 of 30 stocks used to compute the Deutsche Aktienindex (DAX) 67 of 100 stocks used to compute Financial Times Stock Exchange (FTSE 100) and 20 stocks used to compute the Warszawski Indeks Gieldowy (WIG 20). Firstly, in all markets, I have computed correlation coefficients performed on all trading days of the investigated time period from 1.01.1997 to 1.11.2004. According to [3] I have built the Minimal Spanning Tree (MST) for each portfolio of stocks [Fig. 2]. It provides an arrangement of stocks which selects the most relevant connections of each point of the set. In all the markets, the Minimum Spanning Trees visualize economic sectors and subsectors correctly.

On the other hand, companies offering substitute products might be negatively correlated when essentially compete over the same group of customers. Therefore, the success for one company often implies the failure for the others, indicating the market's reaction to the current situation (Futhermore, there is a fundamental negative correlation between gold-related stocks and the rest, indicating the complementary characters of these assets [12]).

I introduce 3 levels of correlations given by (1):

1. Strong (strongly correlated pair of stocks)    $\rho \subset [½, 1]$
2. Weak (weakly correlated pair of stock)         $\rho \subset [0, ½)$
3. Negative (anti-correlated pair of stocks)      $\rho \subset [-1, 0)$

This particular division is necessary to detect how correlation coefficients change in time. Moreover, all levels of correlation coefficients are of importance in portfolio selection [1] and strategy of investment.

In the DJIA portfolio I have detected 9 strongly-correlated pairs (Table 1.1)

**Table 1** Strongly correlated pairs in the DJIA portfolio:

| Pair of stocks | Correlation coefficient |
|---|---|
| C-JPM | 0.72 |
| AXP-C | 0.68 |
| JPM-AXP | 0.65 |
| GE-AXP | 0.61 |
| C-GE | 0.59 |
| JPM-GE | 0.56 |
| IP-DD | 0.5 |
| JNJ-MRK | 0.5 |
| IP-AA | 0.5 |

Other 426 pairs are weakly-correlated. In the DAX portfolio I have detected 205 strongly correlated pairs. The greatest value of correlation coefficient is equal 0,83 (BASF with BAY). I have also detected 1 anti-correlated pair (Allianz AG with DBK). Other 119 pairs are weakly correlated. There is only 1 strongly correlated pair in the WIG portfolio. The correlation coefficient of JLF with KBH is equal 0,50. I have also detected 1 anti-correlated pair (EXB with MEXP). Other 188 pairs are weakly correlated. In the FTSE 100 portfolio I have detected 242 strongly correlated pairs.

## 3 Life time of correlation between stocks prices.

Let us consider the strongly correlated pair Citygroup with JP Morgan (C-JPM) from the DJIA portfolio. Its correlation coefficient performed on all 1969 trading days is constant and equals 0,71. But when the window width $\Delta t$ gets narrowed down, the successive correlation coefficients based even on overlapping data sets might differ from each other (Fig. 1.3). Using a "moving window" $\Delta t$ it is possible to detect how correlation coefficient $\rho$ depends on time.

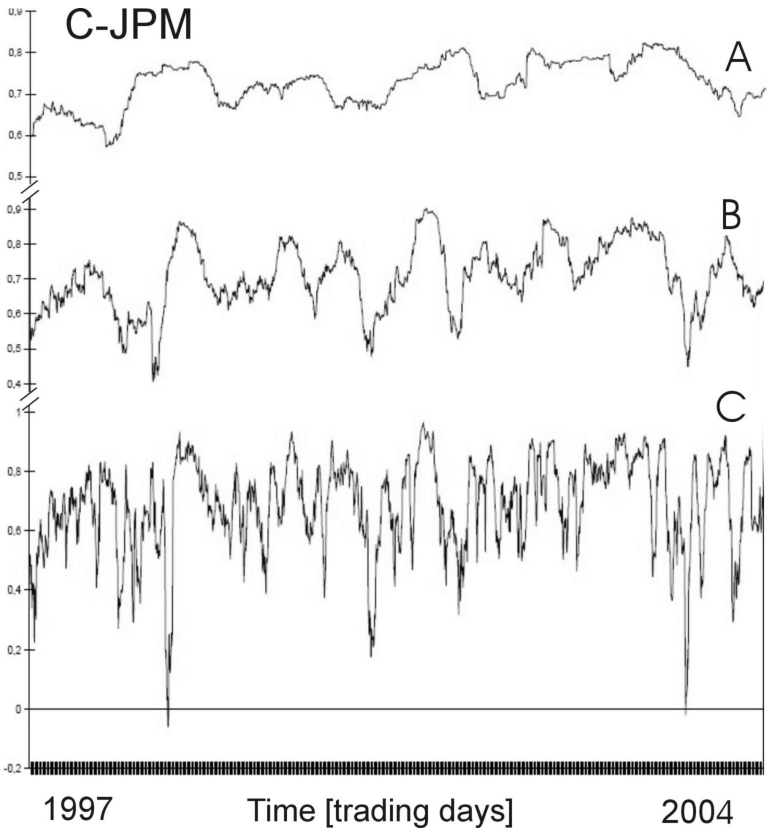

**Fig. 3** The evolution of correlation coefficient $\rho(t, \Delta t)$ between C and JPM obtained for various window width: A) $\Delta t = 200$ trading days [t.d.] B) $\Delta t = 65$ [t.d.] C) $\Delta t = 20$ [t.d.]

For the moving window width $\Delta t=200$ the correlation coefficient $\rho(t)$ between C and JPM changes, but still remains on the strong level. As window width $\Delta t$ is decreasing from 200 to 20 the plots become more noisy. This reflects the fact that market is not static, but dynamic. The underlying price processes are not stationary, so correlations affected by market dynamics change with respect to time because companies and their interactions change. Hence, $\rho(t)$ assumes values from all three levels of correlations: strong, weak and negative.

It is useful to introduce a new tool to measure durability of correlation – the life time $\tau_{ij}(\Delta t)$ of correlation (LTC). It is defined as a length of time when the correlation coefficient $\rho(t,\Delta t)$ is permanently on the strong level. For instance, when $\Delta t$ is equal 65 the correlation coefficient $\rho(t)$ reaches the strong level six times (Fig. 4).

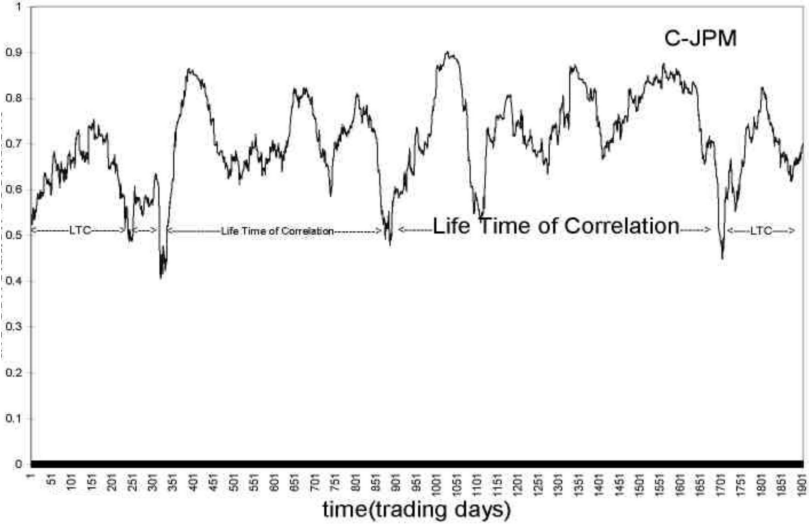

**Fig. 4** The evolution of correlation coefficient $\rho(t, \Delta t)$ between C and JPM obtained for $\Delta t = 65$. The Life-time Of Correlation (LTC) is defined as a lenght of time when the correlation coefficient is permanently on strong level

For each pair it is possible to introduce the mean life time $\langle\tau_{ij}(\Delta t)\rangle$ of correlation (MLTC):

$$\langle \tau_{ij}(\Delta t) \rangle = \frac{1}{N} \sum_{k=1}^{N} \tau_{ijk}(\Delta t) \qquad (3)$$

where $k$ enumerates the successive life times of correlation, $i,j$ are the numerical labels of stocks.

Like $\rho_{ij}(t;\Delta t)$, the MLTC $\langle\tau_{ij}(\Delta t)\rangle$ also depends on the window width $\Delta t$. For instance, when $\Delta t = 65$, the mean life time of correlation $\tau$ C-JPM equals 233,125 trading days.

## 4 The survival and the variety of correlations.

In general, the MLTC is a measure of a durability of correlation performedon all trading days in the investigated time period. I have investigated each of 435 pairs from the DJIA portfolio. For a very small Δt, the correlation $\rho(t)$ is based on on a temporary change of data sets affected by price fluctuations.

For all pairs the mean life times behave in the same way. As Δt increases, $\tau ij(\Delta t)$ grows up because all the plots $\rho(t)$ get smoother (Fig. 3 A). This is an effeect of overlapping data sets that make the correlation change no more.

For strongly correlated pairs, like C-GE from the DJIA portfolio $\tau ij(\Delta t)$ grows up with Δt. On the other hand, weakly correlated pairs, like GM-MO ($\rho_{GM-MO}$ = 0.12) have their lifetime $<\tau_{GM-MO}(\Delta t)>$ = 0 for sufficiently large Δt.

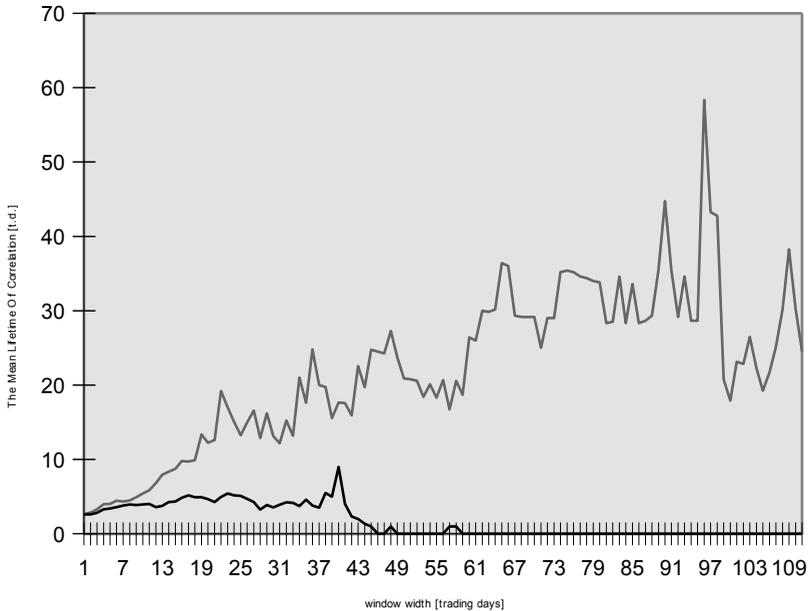

**Fig. 5** The mean life-time $\tau ij(\Delta t)$ of correlation between C-GE (upper line, strongly correlated) and between GM-MO (decaying, weakly correlated) from the DJIA portfolio

On the plot (Fig. 5) the mean life time MLTC is shown for 2 different pairs: C-GE (strongly correlated) and GM-MO (weakly correlated). We see that for the window width $\Delta t$ greater than 60 the mean lifetime $\langle \tau_{GM-MO}(\Delta t) \rangle$ decreases to zero. On the other hand, the mean life time $\langle \tau_{C-GE}(\Delta t) \rangle$ computed for the strongly correlated pair C-GE still grows with $\Delta t$. It is also visible that when $\Delta t$ is too small, the mean life times for all two pairs are small and too similar. So, it is hard to distinguish their levels of correlations. For too large $\Delta t$, however, the MLTC has too large variations. For instance, $\langle \tau_{C-GE}(\Delta t) \rangle$ equals 59 for $\Delta t = 106$, and equals 38 for $\Delta t = 108$. Thus, small increase of $\Delta t$ causes very dramatic change of $\tau(\Delta t)$. This phenomenon has been discovered for all investigated pairs of stocks traded in all markets (DJIA, DAX, WIG, FTSE 100). This observation suggests that the window width shouldn't be too large. The similar results are reported and visualized in Ref. [9] by Onnela et.al. Their results concern the DJIA Minimal Spanning Tree half-life (the time interval in which half of the number of initial connections have decayed).

It is also possible to define the mean life time of correlations MLTC inside a set of stocks used to compute main indices:

$$\langle \tau(\Delta t) \rangle_{IND} = | \frac{1}{N} \sum_{ij} \langle \tau_{ij}(\Delta t) \rangle \qquad (4)$$

where $N$ is a total numer of pairs inside the DJIA, DAX or WIG portfolio. $i,j$ are numerical labels of stocks ($i \neq j$). I introduce this definition to compare the MLTC inside portfoliosand detect the variety of correlations on each of the markets (Fig. 6) . The DJIA and DAX portfolios contain strongly correlated pairs of stocks. It determines the growth of the MLTC with $\Delta t$. Although the MLTC in both portfolios has similar values the DJIA MLTC has large variations. As it was already mentioned, only a few of pairs in the DJIA portfolio are strongly correlated. Their mean life times increases with the window width and then the other pairs mean life times fades. So, the strongly correlated pairs give a contribution to the DJIA MLTC. Therefore, the DJIA MLTC variations are caused by a few strongly correlated pairs. It stands in contrast to the German market where most of the stocks from the DAX portfolio are strongly correlated. Both mean life times of correlations increase and reach 54 when $\Delta t = 90$. On the other hand, the mean life time of correlations inside the WIG portfolio is not bigger than 20. It's a clasic case of differences between emerging (Poland ) and established markets (USA and Germany).

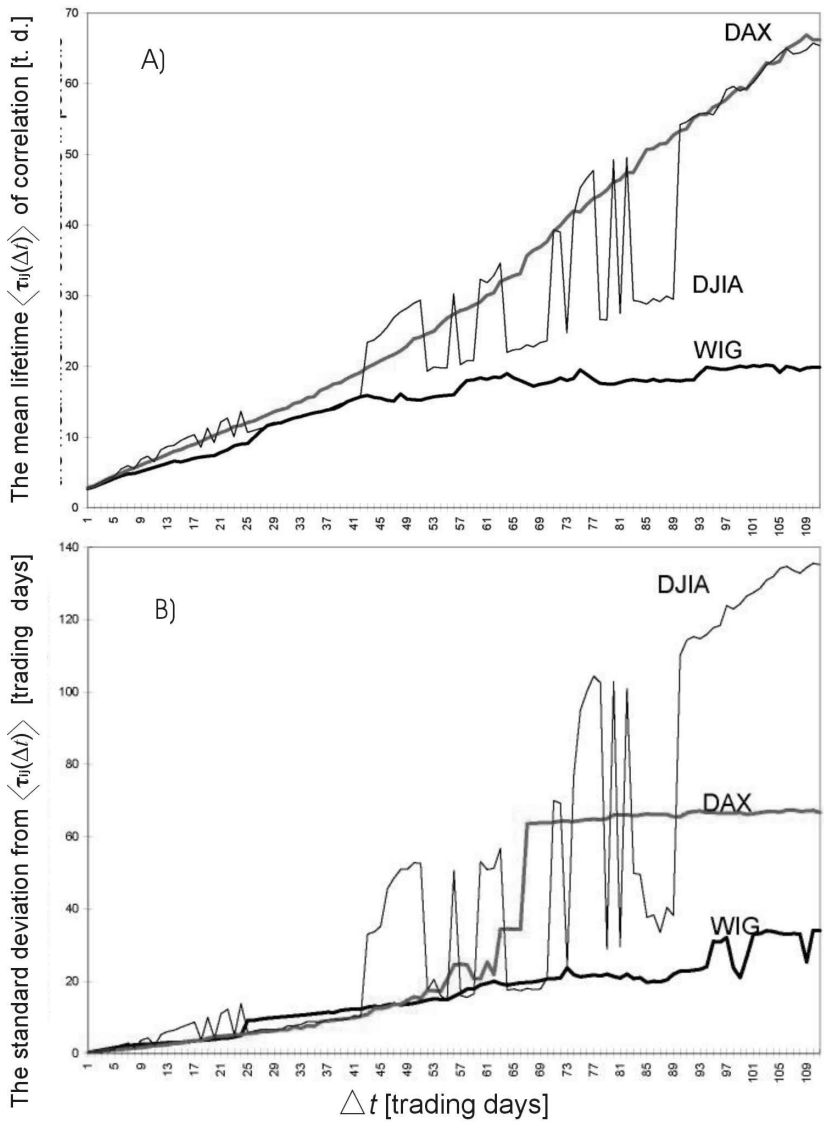

**Fig. 6**
A) The mean life-time <*τij*(Δt)> of correlations inside DJIA, DAX and WIG portfolios
B) The standard deviation from the mean lifetime of correlations inside portfolios

From a practical point of view, it is more useful to analyze the standard optimal window width Δt to detect the variety of correlations between stocks. In all three markets, the standard deviation from MLTC increases with Δt very weakly when the window width is small. Moreover, for Δt = 2, 3, 4, 5 this standard deviation is smaller than 1 in each of the markets!

This situation provides that it is impossible to distinguish the values of correlations between stocks by investigating a very short price history. Like in the DJIA and DAX portfolios, the standard deviation from the <τ(Δt)>$_{WIG}$ grows with Δt. This growth gets weaker when Δt > 60 because we have only one strongly correlated pair (JLF-KBH) in the WIG portfolio. All other pairs are weakly correlated. There is no wide variety. So, the standard deviation from the WIG MLTC increases with the window width weakly.

It is interesting to analyze the standard deviation from <τ (Δt)>$_{DAX}$ for Δt > 65. It stops growing and remains almost constant. It means that the window width Δt = 66 is large enough to detect the variety of correlation in the DAX portfolio.

## 5 Other methods. The Epps Effect

The phenomena of decrease in correlations among price changes in common stocks of companies in one industry for small window width were found in 1977 by Thomas Epps [13]. Although such correlations are not necessarily inconsistent with market efficiency, the data to reveal the presence of lags of an hour or more in the adjustment of stock prices to information relevant to the industry. As the transactions are asynchronous, the correlations measured on short time scale are significantly reduced, which is called the Epps Effect.

From theoretical point of view continuous time random walks [14-15] can be mentioned, that can be used to describe a broad range of processes from transport in disordered solids [16] to finance [17]. Correlated continuous random walks produce asynchronous time series. Cumulating data within a time window suppresses this source of noise but weakens the statistics. The correlation meassure is designed to determinate the grade of co-movement of synchronous observations, while the signals are asynchronous. In order to approach the proper value of the correlation coefficient, the window width Δt should be much larger than the scale of asynchronicity.

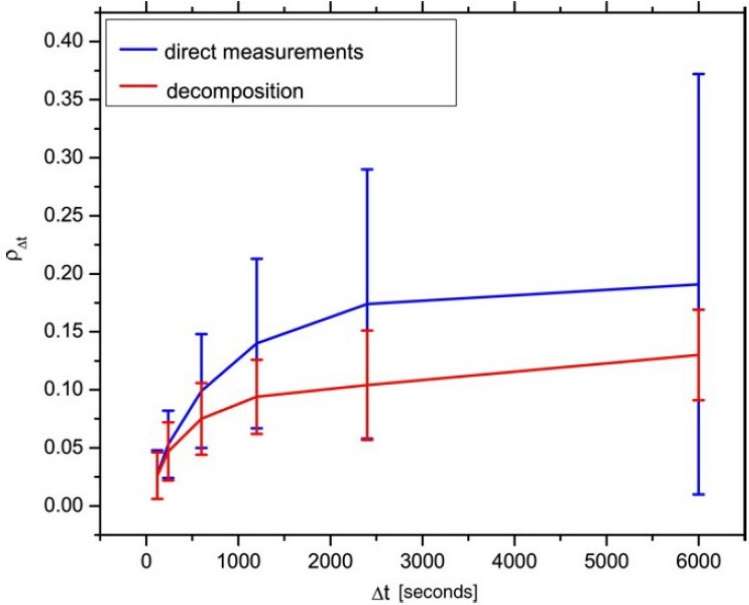

**Fig. 7**. Comparision of the directly measured correlation coefficients and the coefficients determined through the decomposition method in case the real world stock market data. The correlations are computed for the stock pair KO-PEP for the year 2000. The standard deviation of the coefficients obtained through the decompositions are much lower than those of the direct measurements. [1]

However, it is possible to estimate correlations without applying long time windows. In 2009 Janos Kertesz and Bence Toth [18] decomposed the correlations of data cummulated over a long window using a decay of lagged correlations as calculated from short window data. This increased the accuracy of the estimated correlation significantly [Fig. 1.7] and decreased the necessary effort of calculations both in real and computer experiments.

On the other hand, such method of estimation is deficient and questionable, because of economic evolution. There have been attempts using simulation, where the 'true' correlation structure of time series is known because it is fixed beforehand. The estimated correlation coefficients can now be compared against this reference, the 'true' values. But it is risky to claim that in the real world correlations are fixed.

---

1 B.Toth, J.Kertesz, Accurate estimator of correlations between asynchronous signals, Physica A 388 (2009) 1696-1705

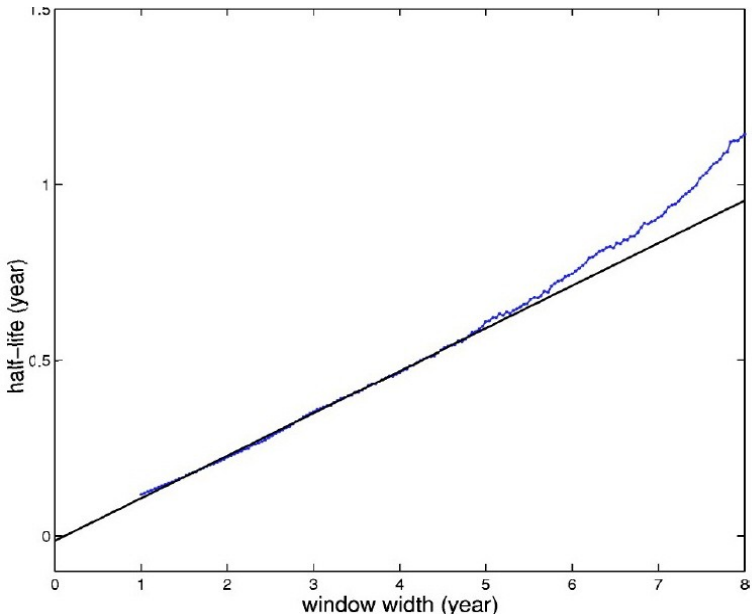

**Fig. 8**. Plot of a S&P tree half-life $t_{1/2}$ as a function of window width $\Delta t$ [2]

Another method, introduced by Onnela et al. [9] refers to Minimum Spanning Trees. It is so called half-life survival ratio t1/2, or tree half-life for short, defined as the time interval in which half of the number of initial connections have decayed. The data set have been obtained for a dynamic asset tree connecting the examined 116 stocks of the S&P 500 index (Fig. 9). The tree has been produced using 4 year window width and centered on January 1, 1998. It seems to have a scale free structure where the scaling exponent of the degree distribution is different for "business as usual" and crash periods. It was found that the stock included in the minimum risk portfolio tend to lie on the outskirts of the asset tree.

The behavior of t1/2 as a function o window width $\Delta t$ is depicted in Fig.1.8. And it is seen to follow a clear linear dependence for $\Delta t$ being between 1 and 5 yr, after which it begins to grow faster than a linear function and behave like the mean lifetime of correlations $\tau_{ij}(\Delta t)$. For the linear region, the tree half-life exhibits t1/2=0,12 $\Delta t$ dependence.

---

[2] J.P. Onnela, A. Chakraborti, K. Kaski, J. Kertesz, A. Kanto, Dynamics Of Market Correlations: Taxonomy And Portfolio Analysis, Physical Review E, 68, 056110, (2003)

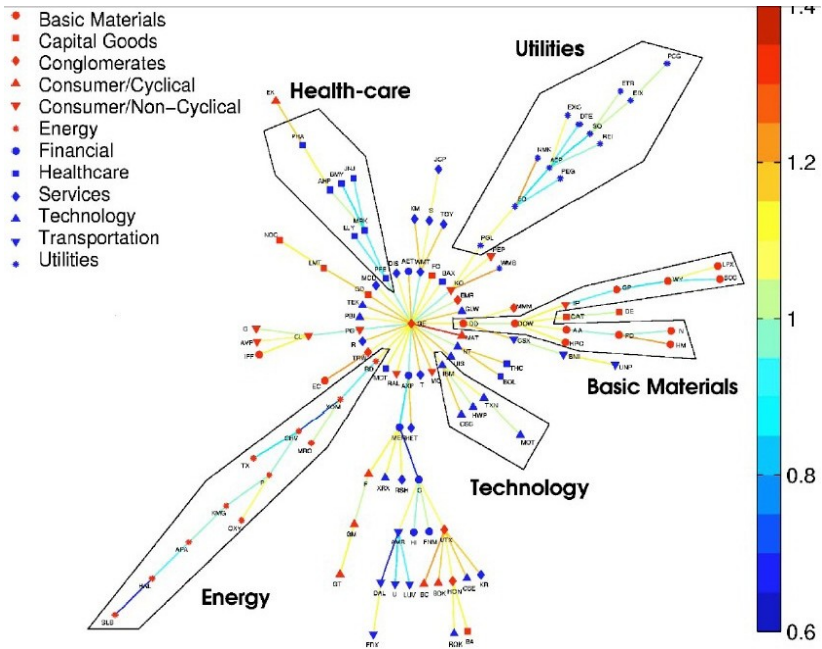

**Fig. 9**. Dynamic asset tree connecting the 116 stocks of the S&P index[3]

# 6 Conclusion

The mean life time of correlation allows to detect differences between durabilities of correlations between stocks traded on established (DJIA, DAX portfolio) and emerging (WIG portfolio) markets. The durability of correlation is determined by the price history and our expectations (the strong level of correlation is separated from weak and negative levels. The mean life time of correlation could measure how the durability of correlation depends on the window width $\Delta t$. According to this, in the investigated markets it is reasonable to choose at least 3 or 4 months from the latest data series to compute correlation coefficients efficiently. In one hand, the window width shouldn't be too small because correlations affected by price fluctuations are hard to distinguish. On the other hand, the window

---

3  J.P. Onnela, A. Chakraborti, K. Kaski, J. Kertesz, A. Kanto, Dynamics Of Market Correlations: Taxonomy And Portfolio Analysis, Physical Review E, 68, 056110, (2003) width shouldn't be too large. Even a small extension of the window width may cause dramatic changes in the mean lifetime of correlation. Some historical data from a distant past are not necessary in building up the optimal durable correlation coefficients. We do not have to know the whole history to make right decisions, especially if some economic connections from the past don't influence on current situations.

In the fifth chapter of this book we show how history influences on hierarchical structure in polish political parties. In other chapters, the natural window width is chosen to be 1 year (Human-human interactions: epidemiology) or one saison (Anomalous interaction in network of Polish football league). Even the phonographic market seems to reveal such properties.

Thus, the knowledge of the life-times offers particular attractions to support managers as it suggests ways by which they adjust their portfolios to benefit from changes in overall market conditions.